\documentclass[a4paper,11pt,twocolumn] {article}

\usepackage[french]{babel}
\usepackage[T1]{fontenc}
\usepackage[latin1]{inputenc}

\usepackage{amsmath} 
\usepackage{amsfonts} 
\usepackage{amsthm} 
\usepackage{epsfig} 
\usepackage{graphicx} 
\usepackage{multirow} 
\usepackage{verbatim} 
\usepackage{supertabular}
\usepackage{color}

\begin{document}

\begin{titlepage}
\title{Analyse du comportement dynamique d'une broche de machine-outils}

\author{Claudiu-Florinel BISU$^{a,1}$, Mihai GHINEA$^{b,1}$,\\
Alain GERARD$^{*c,2}$, Miron ZAPCIU$^{d,1}$, Marin ANICA$^{e,3}$,\\
} 
\date{}

\maketitle

\begin{center}
\textit{
\footnotesize{
$^{1}$ : Université Polytechnique de Bucarest,\\
Laboratoire de Machines et Systèmes de Production\\
Splaiul Independentei 313,   Bucarest - ROUMANIE \\ 
Fax : +(40) 214 104 267\\
$^{2}$ : Université de Bordeaux  - CNRS UMR 5295\\
I2M, Mécanique, Procédés, Interactions\\
351, Cours de la libération, 33405 Talence Cedex - France \\
Fax.: +(33) (0)5 40 00 69 64\\
$^{3}$ : Digitline Company \\
Str. Baneasa, no. 2-6, sector 1, Bucharest, ROUMANIE}
}

\end{center}

\begin{flushleft}
\begin{table*}[h]
	\centering
		\begin{tabular}{lllll}
		E-mail: & claudiu.bisu@upb.ro & & Tél.: & +(40)(0) 724 016 295\\
		E-mail: & ghinea2003@yahoo.com & & Tél.:& +(40)(0) 721 249 578\\
		E-mail: & ajr.gerard@gmail.com & & Tél.:&+(33)(0) 671 707 893\\
		E-mail: &  miron.zapciu@upb.ro & & Tél.:&+(40)(0) 749 206 428\\
		E-mail: & office@digitline.eu  & & Tél.:& +(40)(0) 723 345 015\\
		\end{tabular}
\end{table*}

$^{a}$ : Docteur, Chef de Travaux \\
$^{b}$ : Docteur, Chef de Travaux\\
$^{c}$ : Docteur, Professeur des Universités\\
$^{d}$ : Docteur, Ingénieur, Professeur des Universités\\
$^{e}$ : Ingénieur, Directeur de Digitline\\
\vspace{1.0cm}						
Correspondance à adresser à: \\
Alain GERARD, Professeur Université de Bordeaux et CNRS UMR 5295\\
E-mail: ajr.gerard@gmail.com\\

\end{flushleft}
\end{titlepage}

\newpage
\setlength{\parindent}{0cm}
\textbf{Résumé}
	L'apparition de vibrations, ou bien des régimes instables, est induite par l'interaction dynamique du processus de coupe avec le système élastique de la machine-outil dans différentes conditions de travail. L'objectif de ce travail est de développer un modèle dynamique appliqué au processus de fraisage pour réaliser la surveillance du processus, l'analyse et l'optimisation des conditions de coupe lors du contact outil/pièce/machine. Pour mettre en place ce protocole et atteindre les connaissances approfondies des phénomènes dynamiques nous effectuons une étude de ceux-ci divisée en deux parties : la machine et le processus de coupe. Une analyse dynamique détaillée de la machine est obligatoire avant de modéliser les phénomènes vibratoires présents lors de la coupe.

 \textbf{Mots clés :} vibrations / modèle expérimental / plan des déplacements / vibrations auto-entretenues.

\vspace{1.0cm}	

\textbf{Abstract}  
	 
Vibrations appearance, either unstability system, is the result of the dynamic interaction of the cutting process with the elastic system of machine tools in various working conditions. The objective of this work is to develop a dynamic model applied to process of milling and  to monitor this one, to analyze and to optimize the cutting conditions. To set up this protocol and reach the detailed knowledge of the dynamic phenomena we divide our study into two parts: the machine and the cutting process. A detailed dynamic analysis of the machine is compulsory before modelling the present vibratory phenomena during the cutting.


\setlength{\parindent}{1cm}

\section{Introduction}
\label{sec:Introduction} 

\begin{figure*}[htbp]
\centering
		\includegraphics[width=0.99\textwidth]{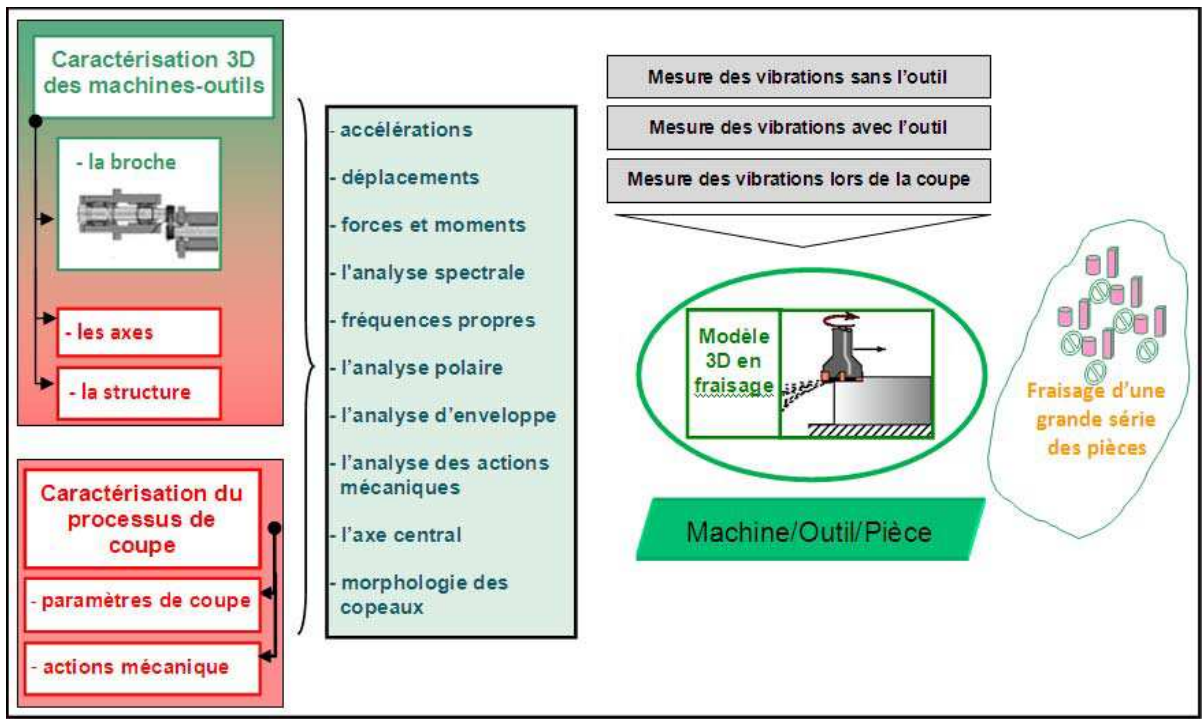}
	\caption{Démarche adoptée.}
	\label{fig1}
\end{figure*}

	Parler d'usinage des pièces en grandes séries, conduit à considérer un ensemble de procédures qui doivent être effectuées par des ingénieurs pour préparer la machine-outil ou le centre d'usinage afin d'atteindre une productivité élevée en termes de qualité et de précision spécifiée par le concepteur \cite{arnaud-AAA-peigne-11}. L'objectif de ce document fait partie d'une recherche plus vaste (Fig.~\ref{fig1}), comportant une analyse expérimentale importante du comportement dynamique de la broche afin de limiter le temps d'intervention et d'entretien lors de l'usinage des grandes séries de pièces ou bien de l'usinage de pièces de grandes dimensions. De plus nous souhaitons obtenir le plus petit nombre possible de rebuts généré par un manque de qualité de surface et de précision géométrique des pièces usinées. En outre, la qualité du comportement dynamique de la broche peut influencer le contact pièce/outil \cite{seguy-AAA-peigne-10}, \cite{kao-lu-07} . Il peut en résulter, implicitement, une influence sur la qualité de la partie active de l'outil tout en gardant néanmoins les coûts de production dans des limites concurrentielles acceptables \cite{{gagnol-AA-barra-06}, {lipski-AAAAA-zaleski-01}}. L'ensemble de la présente recherche est de caractériser le système usinant tridimensionnel (figure~\ref{fig2}), en particulier la broche, les axes linéaires et la structure de la machine-outil. L'objectif est de déterminer les imperfections ou les défauts de fonctionnement dus à l'usure. En effet, ceux-ci peuvent influencer la précision de l'usinage \cite{garnier-furet-00}, \cite{seguy-AAAA-aramendi-08}. Cet article analyse le comportement de la broche en utilisant différentes méthodes d'analyse dynamique. 

Les machines-outils d'aujourd'hui possèdent des comportements dynamiques assez élevés en ce qui concerne les rigidités et les capacités d'amortissement \cite{wehbe-A-arnaud-08}. Cependant les causes d'apparition de vibrations, ou bien des régimes instables \cite{cahuc-AAAAA-gerard-10}, sont données par la dynamique du processus de coupe dans différentes conditions de travail \cite{marui-A-kato-83b}, \cite{arnaud-AAA-peigne-11}. Autrement dit, dans certaines conditions (mauvais choix de paramètres de coupe, frottement intense outil/pièce/copeau, usure de l'outil, faible rigidité de la machine ou de ses composantes), l'interaction du processus de coupe avec le système élastique de la machine-outil provoque l'apparition de vibrations, jusqu'à l'instabilité du processus \cite{tobias-65},
\cite{zapciu-AAA-knevez-09}, \cite{thevenot-AA-larroche-06a}, \cite{thevenot-AA-larroche-06b}. 
Notre objectif est de développer un modèle dynamique tridimensionnel de la coupe en fraisage pour optimiser les conditions de coupe et surveiller le processus de coupe. Pour atteindre la connaissance approfondie des phénomènes dynamiques notre étude dynamique est divisée en deux parties : la machine d'une part et le processus de coupe d'autre part. L'analyse dynamique de la machine est obligatoire avant de pouvoir modéliser les phénomènes vibratoires lors de la coupe \cite{moreau-A-fares-06}.
Dans une première étape nous caractérisons le fonctionnement dynamique de la broche de la machine-outil et nous développons des applications pour l'identification de différents défauts existants dans le comportement dynamique de la broche \cite{zapciu-AAA-knevez-09}. Le but à long terme est d'intégrer tous ces phénomènes et tous ces résultats dans un logiciel spécifique permettant la caractérisation dynamique des machines-outils.
	
\section{Protocole expérimental}
\label{sec:ProtocoleExpérimental}

\begin{figure}[htbp]
	\centering
		\includegraphics[width=0.49\textwidth]{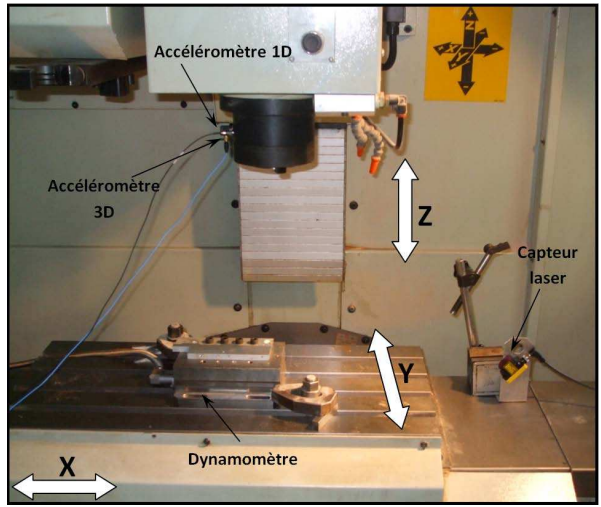}
	\caption{Dispositif expérimental}
	\label{fig2}
	\end{figure}

	Un protocole expérimental est mis en place, pour obtenir les informations nécessaires à l'analyse des phénomènes provenant de la machine ou bien de la coupe. Le but des essais est d'identifier le comportement vibratoire de la machine-outil en utilisant une méthode complète de caractérisation. Les mesures sont effectuées sur un centre d'usinage vertical trois axes à commande numérique, muni d'une broche de 11 kW. La vitesse de rotation maximale est de 8.000 tr/min (Fig.~\ref{fig2}).

Le comportement dynamique est identifié par un accéléromètre tridimensionnel et par un accéléromètre unidirectionnel. Ces appareils sont fixés sur le corps de la broche. Par la suite l'accéléromètre tridimensionnel est monté sur la pièce pour mettre en correspondance le processus de coupe et le comportement de la broche. L'acquisition et l'analyse des signaux sont réalisées grâce à l'équipement \textit{Digitline- Fastview} développé au sein du laboratoire de \textit{Machine et Système de Production de l'Université Politehnica de Bucarest.}

Les efforts sont mesurés à l'aide d'un dynamomètre à trois composantes, type Kystler 9257B, tandis que l'évolution de la vitesse instantanée de l'outil est donnée par un capteur rotatif. Le caractère dynamique tridimensionnel est mis en évidence en cherchant les différentes corrélations existantes ou les différentes évolutions des paramètres qui permettent de caractériser la dynamique de la broche \cite{bisu-AAA-minciu-10}. Cette analyse fournit les informations nécessaires à la conception et au développement d'un modèle expérimental de caractérisation dynamique d'une machine-outil en prenant en compte les influences dynamiques lors du contact broche-outil/coupe ou bien les influences du processus de coupe sur l'outil respectivement sur la broche.

\begin{table*}[htbp]
	\centering
	\caption{Paramètres de coupe.} 
	\begin{tabular}{c c c c c c} 
\hline
&  &  &  &  & \\
Fraise& N & f & $V_{f}$ & $V_{c}$ & $a_{p}$ \\

 & $(tr/min)$ & $(mm/dent)$ & $(mm/min)$ & $ m/min)$ & (mm) \\
 \hline
 \multirow{4}* & 1.000 & 0,066 & 396 & 251,32 &   \\
{diamètre 80 mm} & 2.000 & 0,033 & 396 & 502,64 & 1 \\
{ (z = 6 dents)}& 5.000 & 0,033 & 990 & 1.256,60 & \\
&  &  &  &  & \\
\hline

\end{tabular}
	\label{tabl-1}
\end{table*}

Les essais sont réalisés en deux étapes : fonctionnement de la machine libre (sans processus de coupe) d'une part et d'autre part fonctionnement de la machine en charge (avec le processus de coupe activé). Dans le fonctionnement à vide on augmente (resp. diminue) la vitesse de rotation par paliers de 500 tr/min jusqu'à atteindre 8.000 tr/min (resp. 500 tr/min). Les conditions d'essais de la machine en charge (pendant la coupe) sont présentées dans le tableau\ref{tabl-1}.

\section{Etude du comportement dynamique de la broche}
\label{sec:EtudeDeComportementDynamiqueDeLaBroche}

La campagne d'essais est conçue dans le but mettre en évidence le comportement dynamique de la broche en analysant chaque influence qui peut générer des vibrations ou des instabilités \cite{bisu-AA-knevez-09}. L'étude est développée dans trois configurations : la première consiste en l'analyse de la broche libre, sans l'outil, la deuxième avec l'outil et la troisième configuration pendant la coupe.

\subsection{Etude dynamique de la broche libre}
\label{sec:EtudeDynamiqueDeLaBrocheLibre}

\begin{figure}[htbp]
	\centering
		\includegraphics[width=0.49\textwidth]{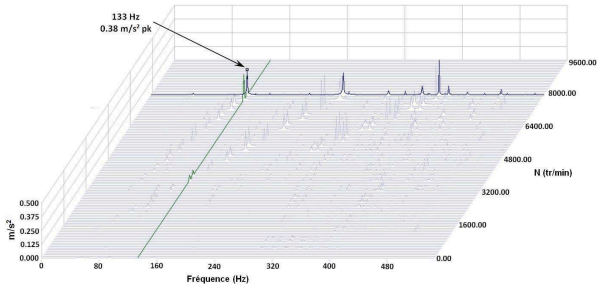}
	\caption{Diagramme Waterfall des  accélérations pour la variation des vitesses de 500 tr/min à 8.000 tr/min et de 8.000 tr/min à 500 tr/min, broche libre sans l'outil.}
	\label{fig3}
\end{figure}

Nous mesurons d'abord la vitesse de rotation de la broche sans l'outil, puis avec l'outil et en fin pendant le processus de coupe. Le premier test effectué, Figure~\ref{fig3} (mesures dans la direction d'avance X), montre l'accroissement de l'amplitude des vibrations en fonction de la vitesse de coupe, notamment à partir de 5.000 tr/min et jusqu'à 8.000 tr/min. L'analyse des accélérations est effectuée pour identifier les différents défauts existants sur la chaîne cinématique, mais aussi en vitesse de vibration et de déplacement.

\begin{figure}[htbp]
	\centering
	\includegraphics[width=0.49\textwidth]{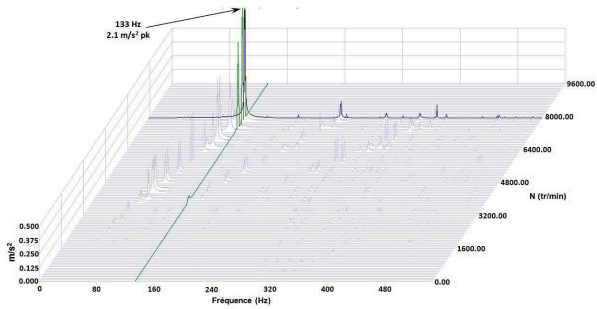}
	\caption{Spectre des fréquences pour la vitesse de rotation de 8.000 tr/min sur les trois directions X, Y et Z de la machine.}
	\label{fig4}
\end{figure}

Nous observons une augmentation de l'amplitude des vibrations. Nous analysons le spectre de fréquence pour chaque palier de vitesse de rotation. Les fréquences génératrices des vibrations sont alors recherchées. La Figure~\ref{fig4} présente l'étude des fréquences données par l'accéléromètre sur les trois directions de mouvement de la machine, plus spécialement dans les conditions de vitesse à 8.000 tr/min. Sur la fréquence fondamentale (première harmonique qui correspond à la fréquence de rotation) nous trouvons une amplitude très importante par rapport à l'amplitude globale. Par exemple, sur X nous avons une amplitude de 0,52 mm/s pour la vibration globale et de 0,45 mm/s pour l'harmonique du premier ordre. De même sur Y nous avons 0,98 mm/s pour la vibration globale et 0,97 mm/s pour l'harmonique du premier ordre. Enfin, sur Z nous constatons 0,3 mm/s en vibration globale et 0,19 mm/s d'amplitude sur l'harmonique du premier ordre. Lors de cette analyse pour chaque vitesse de rotation utilisée nous constatons la persistance d'une amplitude élevée de la fréquence fondamentale donnée par la vitesse de rotation de la broche. L'équilibrage de la broche s'avère donc nécessaire pour un bon fonctionnement de celle-ci. A présent, nous nous intéressons au fraisage des pièces de grandes dimensions. Dans cette optique, nous utilisons des fraises de grand diamètre. L'analyse dynamique est menée sur le comportement de la broche munie d'une fraise neuve de diamètre 80 mm type Sandvik Coromant, à six dents (R365-080Q27-15M). Lors des essais effectués avec l'outil, en suivant le même protocole expérimental que précédemment, nous déterminons le diagramme de Waterfall présenté à la Figure~\ref{fig5}. Nous observons que l'amplitude atteinte est de 2,05m/s$^{2}$ pour la vitesse de rotation de 8.000tr/min. L'amplitude présente le même comportement de croissance en fonction de la vitesse. Cependant, la croissance de l'amplitude des vibrations correspondant à la fréquence fondamentale est plus importante avec l'outil que sans l'outil.

\subsection{Etude dynamique de la broche avec l'outil}
\label{sec:EtudeDynamiqueDeLaBrocheAvecLOutil}

\begin{figure}[htbp!]
	\centering
		\includegraphics[width=0.49\textwidth]{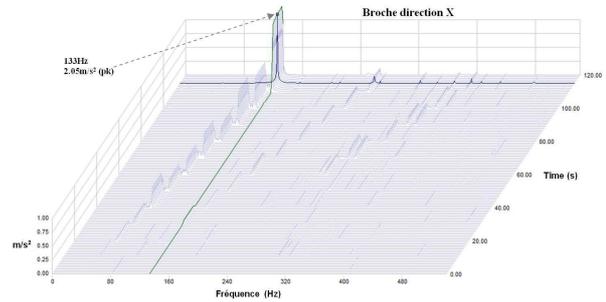}
	\caption{Diagramme de Waterfall des accélérations pour la variation de vitesse de 500 tr/min à 8.000 tr/min, broche avec l'outil, fraise de diamètre 80 mm à 6 dents.}
	\label{fig5}
\end{figure}

Maintenant l'analyse est approfondie au niveau de l'outil pour caractériser son état dynamique et son évolution en fonction de la vitesse de rotation \cite{namazi-AA-rajapakse-07}. La nécessité de développer des applications à l'aide du traitement du signal nous est imposée par les exigences de qualité des produits manufacturés, de maintenance et de coûts de production. L'analyse des signaux est effectuée pour 98.304 lignes spectrales. Ensuite nous utilisons une analyse FFT synchrone et par une double intégration nous obtenons les déplacements. Comme l'équilibrage peut avoir des influences vibratoires sur le comportement dynamique de l'outil \cite{bissey-A-lapujoulade-05} nous nous efforçons de déterminer les différents effets des vibrations sur celui-ci. Nous déterminons l'ellipse des déplacements de vibration dans le plan (X, Y) et le diagramme polaire pour les composantes des vecteurs de vibrations selon X, Y et Z. Nous privilégions deux configurations de la position de l'outil dans le porte-outil : à 0$^{\circ}$ (première position de l'outil dans le cône de la broche) et à 180$^{\circ}$ (deuxième position de l'outil tourné de 180° par rapport à la première position). Celles-ci correspondent en effet à l'harmonique du premier ordre, Figure~\ref{fig5}.

\begin{figure}[htbp!]
	\centering
		\includegraphics[width=0.49\textwidth]{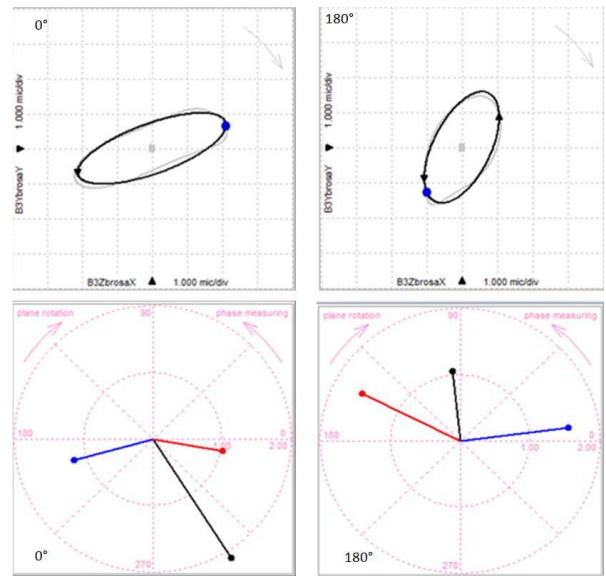}
	\caption{Diagrammes orbital et polaire des déplacements dans deux configurations de montages de l'outil dans le porte-outil à 0$^{\circ}$ et à 180$^{\circ}$.}
	\label{fig6}
\end{figure}

Dans la position à 0$^{\circ}$ nous déterminons les caractéristiques de l'ellipse des déplacements. Son grand axe est de 4,5 $µm$ et le petit axe de 1,4 $µm$. L'angle d'inclinaison du plan de l'ellipse est de 20$^{\circ}$. De même pour la position à 180$^{\circ}$ l'angle d'inclinaison du plan de l'ellipse est de 63$^{\circ}$ tandis que le grand axe est de 3,4 $µm$ et le petit axe de 1,6 $µm$ Figure~\ref{fig6}. Avec ces différentes amplitudes et grâce au diagramme polaire nous obtenons les amplitudes et la position angulaire pour les trois directions. Nous remarquons une déficience de l'outil.

\subsection{Etude dynamique de la broche lors de la coupe}
\label{sec:EtudeDynamiqueDeLaBrocheLorsDeLaCoupe}

L'étude dynamique continue avec l'analyse du comportement pendant la coupe. Pour ces essais, l'outil utilisé est de type Sandvik R365-080Q27-15M (plaquette de carbure CBN). Le matériau usiné est de type aluminium. Pour chaque essai, en fonction de la vitesse de coupe et d'avance, les efforts de coupe sont mesurés en même temps que les vibrations.

\begin{figure*}[htbp]
	\centering
		\includegraphics[width=0.99\textwidth]{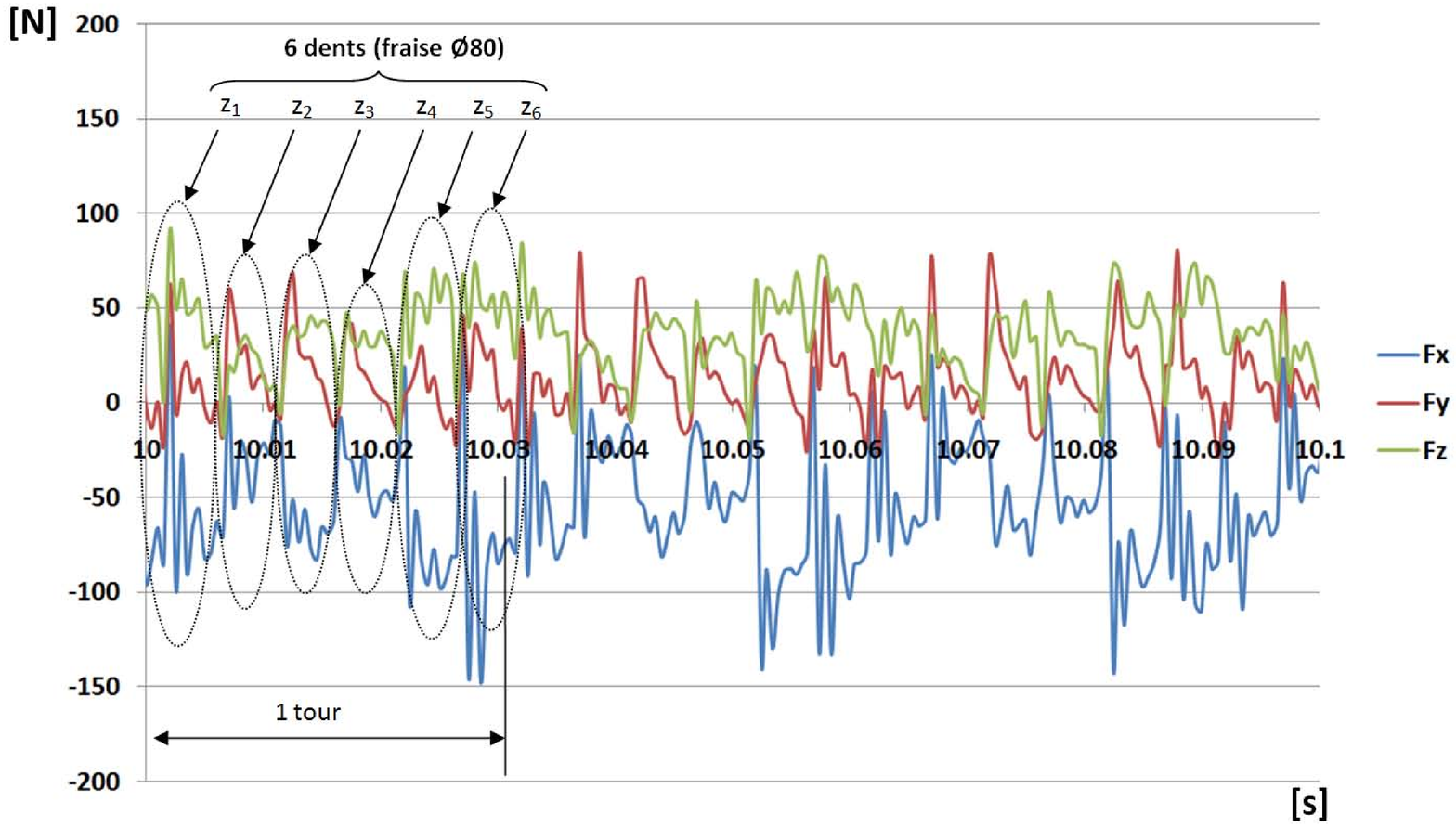}
	\caption{Signaux des composantes de la résultante sur les trois directions de coupe pour trois tours dans le cas : ap= 1 mm, f= 0,033 mm/dent et N = 2.000 tr/min.}
	\label{fig7}
\end{figure*}

\begin{figure}[htbp!]
	\centering
		\includegraphics[width=0.49\textwidth]{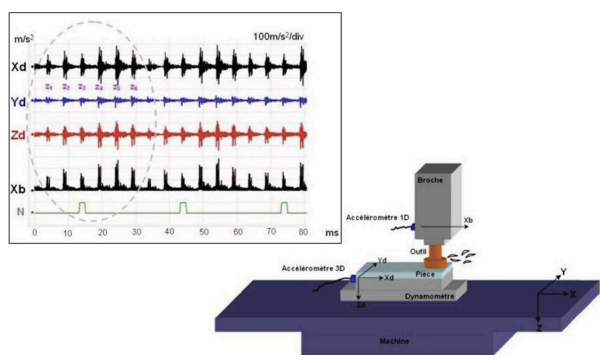}
	\caption{Mesures des vibrations et position de l'accéléromètre 3D et de l'accéléromètre 1D pendant la coupe}
	\label{fig8}
\end{figure}

Lors des mesures effectuées nous observons qu'il y a des dents qui sont plus chargées que d'autres Figure~\ref{fig7}. Les résultats dynamométriques obtenus sont validés par les signaux (Xd, Yd et Zd) obtenu grâce à l'accéléromètre tridimensionnel, fixé sur la pièce, et aussi grâce à l'accélération mesurée sur le corps de la broche, Xb dans la direction X de la machine (Fig.~\ref{fig8}). Le caractère variable du travail des dents pendant la coupe a une influence importante sur le comportement dynamique de la broche que nous pouvons analyser à partir des données accélérométriques mesurées par les accéléromètres.

\subsection{Etude dynamique par l'enveloppe dymanique}
\label{sec:EtudeDynamiqueParLEnveloppeDynamique}

\begin{figure}[htbp]
\centering
		\includegraphics[width=0.48\textwidth]{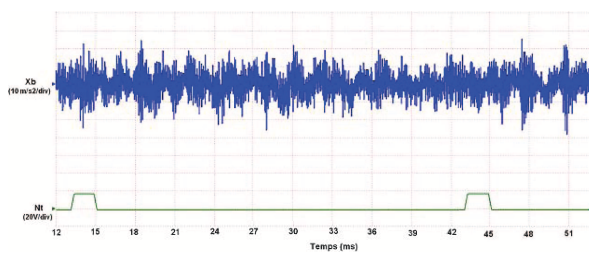}
	\caption{Signal des vibration mesuré dans la direction X de la broche.}
	\label{fig7b}
\end{figure}

L'analyse dynamique est concentrée maintenant sur la caractérisation de la broche lors de la coupe pour identifier les défauts possibles existant sur les éléments de la broche. Nous mesurons le signal des vibrations en accélération Xb obtenu grâce à l'accéléromètre unidirectionnel positionné sur le corps de la broche (dans la direction X, du coté des paliers avant et arrière) et synchronisé avec le signal de la vitesse de rotation (Nt) Figure~\ref{fig7b}. 

\begin{figure}[htbp]
	\centering
		\includegraphics[width=0.47\textwidth]{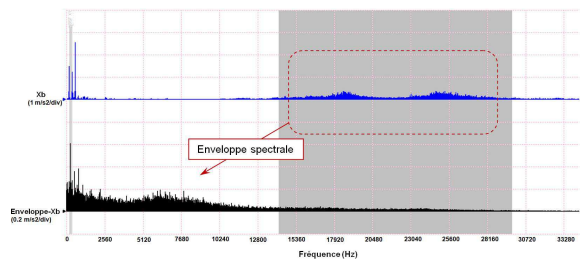}
	\caption{Spectre des fréquences et zone de fréquence identifiée par l'enveloppe.}
	\label{fig7c}
\end{figure}

Par l'analyse FFT nous avons trouvé les fréquences données par la coupe et la plage des fréquences de résonance (ou bien d'excitation) sur laquelle nous allons appliquer la méthode de l'enveloppe pour vérifier l'existence des différents défauts sur les paliers de la broche. Avec le spectre des fréquences lors de la coupe nous pouvons analyser les fréquences en deux étapes : les fréquences de travail de la fraise en coupe et les fréquences de résonance encadrées (Fig.~\ref{fig7c}) pour analyser le comportement dynamique des roulements de la broche.

L'analyse de l'accélération sur la broche dans la direction X montre l'existence lors de FFT des amplitudes élevées à haute fréquence (en bleu à gauche Figure~\ref{fig7c}). En faisant appel à la méthode d'analyse en fréquence par l'enveloppe des vibrations nous déterminons la plage des fréquences donnée par l'excitation des paliers et donc la possibilité d'avoir des défauts sur les roulements de la broche.

\begin{figure}[htbp]
	\centering
		\includegraphics[width=0.47\textwidth]{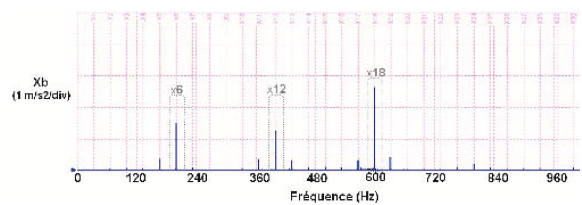}
	\caption{Spectre de fréquence lors de la coupe dans la direction X de la broche.}
	\label{fig9b}
\end{figure}

Avant de lancer l'analyse par enveloppe nous analysons le processus de coupe et dans la Figure~\ref{fig9b} nous observons les fréquences de travail de la fraise, ayant l'harmonique d'orde 6, équivalent au six dents de la fraise. Cette fréquence peut être considérée comme la fréquence de travail des dents de la fraise dans la matière. En suite nous remarquons la présence des harmoniques d'ordre x12 et x18 correspondant respectivement aux multiples d'ordre 2 et 3 de la fréquence d'ordre 6. Cette analyse met en évidence le comportement dynamique de la fraise pendant la coupe, et fournit de très importantes informations sur la qualité de travail des dents.

Pour identifier les différents défauts existants sur les paliers de la broche nous utilisons la méthode d'enveloppe basée sur les particularités de construction des roulements et mettons en évidence la présence des chocs et des frottements de ces éléments. Contrairement à d'autres sources de vibrations dues au taux élevé de variation des forces, les chocs peuvent exciter le corps du palier sur sa fréquence de résonance. L'intérêt dans le diagnostic des roulements est la fréquence d'apparition et l'amplitude de ces oscillations. Bien que la fréquence de résonance se manifeste dans une plage de fréquence relativement étroite, la modulation en amplitude due à ces chocs et la variation de conditions de transmission, nécessitent une analyse dans une large plage de fréquence, centrée sur la fréquence de résonance. La croissance de la plage d'analyse est souvent limitée par l'existence de signaux de haute fréquence qui se superposent sur cette bande.

 Les fréquences d'apparition des chocs engendrées par les défauts de roulement portant dépendent de la vitesse et la géométrie du roulement concerné. Compte tenu de la vitesse de rotation mesurée il résulte une relation directe entre la fréquence mesurée des défauts et le type de défaut de roulement. Il y a cinq fréquences caractéristiques pour lesquelles nous pouvons déterminer le type de défaut précisément. La fréquence de rotation de l'arbre $f_{n}$, la fréquence fondamentale de la cage $F_{C}$ (~\ref{Fc}), la fréquence de passage de billes sur l'anneau intérieur BPFI (Equ.~\ref{BPFI}), la fréquence de passage des billes sur l'anneau intérieur BPFO (~\ref{BPFO}), la fréquence de rotation des billes BSF (~\ref{BSF}), et la fréquence de défaut local de la bille BFF (~\ref{BFF}), \cite{mendel-aa-batista-09}, \cite{mobley-99}:

 \begin{equation}
			\label{Fc}
			F_{C} =\frac{1}{2}f_{n}\left(1-\frac{D_{b}\cos \theta}{D_{c}}\right),
		\end{equation}

  \begin{equation}
			\label{BPFI}
			 BPFI =\frac{N_{b}}{2}f_{n}\left(1+\frac{D_{b}\cos \theta}{D_{c}}\right),
		\end{equation}

  \begin{equation}
			\label{BPFO}
		 	BPFO =\frac{N_{b}}{2}f_{n}\left(1-\frac{D_{b}\cos \theta}{D_{c}}\right),
		\end{equation}

 \begin{equation}
			\label{BSF}
		 BSF =\frac{D_{c}}{2D_{b}}f_{n}\left(1- \frac{D^{2}_{b}\cos^{2} \theta}{D^{2}_{c}}\right),
		\end{equation}
 
\begin{equation}
			\label{BFF}
		 BFF =\frac{D_{c}}{D_{b}}f_{n}\left(1- \frac{D^{2}_{b}\cos^{2} \theta}{D^{2}_{c}}\right),
		\end{equation}

 où $D_{b}$ représente le diamètre de la bille, $\theta$ est l'angle de contact base sur le ratio entre l'effort axial et l'effort radial, $D_{c}$ représente le diamètre de la cage et $N_{b}$ le nombre de billes.
 L'enveloppe du signal des vibrations présente un signal basse fréquence qui suit les pics du signal d'entrée lors de la transformée d'Hilbert. Dans le spectre des fréquences de l'enveloppe on trouve les composantes des fréquences égales au taux d'apparition des impacts et une amplitude proportionnelle à leur énergie.

 Dans une deuxième étape, connaissant la série de roulement (7012C), nous pouvons déterminer les fréquences des défauts de chaque composant du roulement. 

\begin{figure}[htbp]
	\centering
		\includegraphics[width=0.48\textwidth]{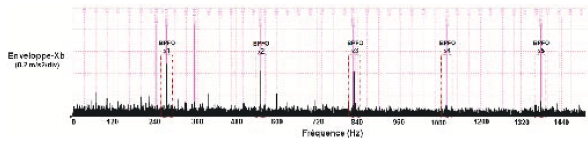}
	\caption{Spectre de fréquence avec la méthode de l'envelope spectrale}
	\label{fig9c}
\end{figure}

Les fréquences de défauts sont calculées par les données concernant la géométrie de roulement \cite{bisu-AAA-anica-10}. Connaissant la série de roulement avant et arrière nous identifions les fréquences correspondantes à partir de l'enveloppe spectrale. Pour accroître la qualité de l'analyse, nous effectuons la transformée FFT synchrone et nous éliminons les composantes spectrales générées par la coupe. Ensuite par filtrage des fréquences de résonance et par application de la transformée de Hilbert suivie de la transformée FFT nous obtenons le diagramme de la Figure~\ref{fig9c}. Nous remarquons l'existence d'une fréquence de défaut sur le roulement avant (7012C) correspondant à la bague extérieure du roulement (BPFO), visible aussi sur les harmoniques de la fréquence de défaut (Fig.~\ref{fig9c}). Par le traitement du signal de l'enveloppe spectrale nous pouvons introduire le marqueur correspondant au défaut et après nous identifions la fréquence et ses harmoniques dans l'enveloppe spectrale.

\section{Conclusions}
\label{sec:Conclusions}

Le but de cet article était de présenter des éléments nécessaires à l'identification du comportement dynamique d'une machine-outils. Nous avons développé une méthode d'analyse pour la caractérisation dynamique appliquée à une broche de machine-outils et aussi pour l'ensemble broche-outil-processus de coupe en obtenant des informations qualitatives sur les paliers de la broche et l'outil. L'application de cette méthode est indispensable tant au niveau de la qualité des surfaces usinées que de la surveillance et de la maintenance. Par la suite nous pensons développer la méthode de l'enveloppe spectrale pour caractériser le comportement dynamique de la fraise pendant la coupe afin d'optimiser les différents paramtères d'usinage. Enfin notre recherche est orientée vers un modèle dynamique tridimensionnel prenant en compte la géométrie de l'outil et les différents phénomènes dynamique lors de la coupe, modélisation nécessaire pour une caractérisation qualitative du comportement de la machine-outil.

\section*{Remerciements}
\label{sec:Remerciements}

Ce travail a été soutenu par CNCSIS-UEFISCU, projet PNII-RU, code 194/2010


\end{document}